\documentclass[10pt]{article}
\usepackage[latin9]{inputenc}
\usepackage{geometry}
\usepackage{color}
\usepackage{textcomp}
\usepackage{amsmath}
\usepackage{amssymb}
\usepackage{graphicx}
\PassOptionsToPackage{normalem}{ulem}
\usepackage{ulem}
\usepackage[unicode=true,
 bookmarks=false,
 breaklinks=false,pdfborder={0 0 1},backref=false,colorlinks=false]
 {hyperref}

\makeatletter

\newcommand{\lyxmathsym}[1]{\ifmmode\begingroup\def\b@ld{bold}
  \text{\ifx\math@version\b@ld\bfseries\fi#1}\endgroup\else#1\fi}

\@ifundefined{date}{}{\date{}}
\usepackage{cite,times}

\makeatother

\begin{document}
\title{Optical designs for realization of
a set of schemes for quantum cryptography}
\author{{\normalsize{}Mitali Sisodia$^{\spadesuit}$, Kishore Thapliyal$^{\ddagger,}$}\thanks{Email: kishore.thapliyal@upol.cz }{\normalsize{},
and Anirban Pathak$^{\spadesuit,}$}\thanks{Email: anirban.pathak@gmail.com}{\normalsize{}
}\\
{\normalsize{}$^{\spadesuit}$Jaypee Institute of Information Technology,
A-10, Sector-62, Noida, UP-201309, India}\\
{\normalsize{}$^{\ddagger}$RCPTM, Joint Laboratory of Optics of Palacky
University and Institute of Physics of Academy of Science }\\
{\normalsize{}of the Czech Republic, Faculty of Science, Palacky University,
17. listopadu 12, 771 46 Olomouc, Czech Republic}}
\maketitle
\begin{abstract}
Several quantum cryptographic schemes have been proposed and realized
experimentally in the past. However, even with an advancement in quantum
technology and escalated interest in the designing of direct secure
quantum communication schemes there are not many experimental implementations
of these cryptographic schemes. In this paper, we have provided a
set of optical circuits for such quantum cryptographic schemes, which
have not yet been realized experimentally by modifying some of our
theoretically proposed secure communication schemes. Specifically,
we have proposed optical designs for the implementation of two single
photon and one entangled state based controlled quantum dialogue schemes
and subsequently reduced our optical designs to yield simpler designs
for realizing other secure quantum communication tasks, i.e., controlled
deterministic secure quantum communication, quantum dialogue, quantum
secure direct communication, quantum key agreement, and quantum key
distribution. We have further proposed an optical design for an entanglement
swapping based deterministic secure quantum communication and its
controlled counterpart. 
\end{abstract}

\section{Introduction\label{sec:Introduction}}

\textcolor{black}{The first protocol for quantum key distribution
(QKD) was proposed by Bennett and Brassard \cite{bennett1984quantum},
which provided security not conditioned on computationally complex
tasks. Since then it has been strongly established that the quantum
cryptography can provide unconditi}onal security \cite{shor2000simple,renner2008security},
which is a clear advantage over its classical counterparts. Inspired
by this advantage, various schemes of QKD and other cryptographic
tasks have been proposed (\cite{gisin2002quantum,pirandola2019advances,pathak2013elements}
and references therein). Some of them have also been realized experimentally
(\cite{bennett1992experimental,lo2014secure} and references therein).
Interes\textcolor{black}{tingly, the quantum resources and the experimental
techniques used in these successful experiments are not the same.
To stress on this particular point, we may note that the QKD entered
into the experimental era with the pioneering experimental work of
Bennett et al. \cite{bennett1992experimental}. In this work, randomly
prepared polarization states of single photons were used; but as there
does not exist any on-demand single photon source, they used faint
light beams generated from light emitting diode as approximate single
photon source. If such a source is used, to circumvent photon number
splitting (PNS) attack \cite{brassard2000limitations}, one has to
use decoy qubits to obtain security \cite{lo2005decoy}. Obviously,
in 1992 work of Bennett et al., no decoy qubit was used. However,
in later experiments, decoy states have been frequently used. For
example, in 2006, Zhao et al., had realized a decoy state based QKD
protocol \cite{zhao2006experimental} using the acousto-optic modulators
(AOMs) to achieve polarization insensitive modulation. In the absence
of on-demand single photon sources, various strategies have been used
to realize single photon based QKD schemes, like BB84 \cite{bennett1984quantum}
and B92 \cite{bennett1992quantum}. Some of the experimentalists have
used weak coherent pulse (WCP) as an approximate single photon source
\cite{diamanti2016practical,duplinskiy2017low,kiktenko2017demonstration,koashi2006efficient,korzh2015provably,lo2014secure,wang2016experimental,xu2015experimental,gleim2016secure},
while the others have used heralded single photon source \cite{soujaeff2007quantum,wang2008experimental}.
Qubits are al}so realized in the polarization \cite{bennett1992experimental},
phase \cite{zhao2006experimental}, time-bin \cite{gisin2003quantum},
and frequency \cite{sun1995long} encodings of the p\textcolor{black}{hotons
(see \cite{gisin2002quantum} for review). In the last 28 years, a
continuous progress has been observed in the experimental QKD. It
started from the experimental realization of a single photon based
QKD scheme using WCP, but as time passes many other facets of QKD
have been experimentally realized. For example, in one hand MDI-QKD
has been realized using untrusted source \cite{lo2012measurement,liu2016polarization}
and heralded single photon source \cite{zhang2018biased,zhou2016measurement}.
On the other hand, soon it was realized that continuous variable QKD
can be used to circumvent the need of single photon sources and thus
to avoid several attacks (\cite{srikara2019continuous} and references
therein). Naturally, some of the continuous variable QKD schemes have
been realized in the recent past \cite{grosshans2002continuous,gottesman2001secure,ralph1999continuous}.
}Alternative schemes based on distributed phase reference QKD are
also proposed and realized, for example, we may mention differential
phase shift \cite{takesue2007quantum} and coherent one-way \cite{stucki2009high}
protocols. It is also worth mentioning about the schemes for counterfactual
quantum cryptography \cite{brida2012experimental}, QKD with orthogonal
states \cite{avella2010experimental} and higher dimensional quantum
systems \cite{groblacher2006experimental}, and secure quantum communication
with a classical party \cite{boyer2017experimentally}.  Beyond this,
to address the concerns of the end users, over the time the devices
used by them have become portable (say, a silicon photonic transmitter
is designed for polarization-encoded QKD \cite{ma2016silicon,ding2017high}
and chip-based QKD systems h\textcolor{black}{ave been realized \cite{sibson2017chip});
QKD has been realized using erroneous source \cite{xu2015experimental};
key generation rate over noisy channel has been increased (e.g., in
\cite{zhang2018experimental}, a key generation rate of 1.3 Gbit/s
was achieved over a 10-dB-loss channel); distance over which a key
can be securely distributed has been increased, for example, in \cite{boaron2018secure},
QKD is performed over 421 km in optical fiber, underwarter QKD is
reported over 23 km optical fiber \cite{muller1995underwater}, and
in the last 2-3 years, a couple of QKD experiments have been performed
using satellites \cite{khan2018satellite,liao2018satellite} - the
one which needs special mention is the quantum communication between
the ground stations located at China and Austria at a distance of
about 7600 km \cite{liao2018satellite}. Furthermore, various commercial
products, like Clavis 2 and Clavis 3 of ID Quantique \cite{IDQ} and
MagiQ QPN of MagiQ \cite{mgQ}, have also been marketed. }
Clavis 2 was hacked later that is why it is not available now\textcolor{black}{{}
Experimental implementations of secure communication schemes have
not only revealed several interesting features relevant for quantum
foundations, but have also led to new technologies having clear advantages
over the traditional technologies. For instance, plug and play QKD
systems to circumvent phase and polarization compensation required
in optical fibers. Similarly, an eavesdropper may design specific
attacks depending upon the configuration--Trojan-horse attack \cite{gisin2006trojan}
which can be circumvented by clever use of technology. Additionally,
eavesdropping attempts exploiting side-channels or device imperfection
also depend on the implementation (see \cite{pirandola2019advances}
for review). Aim of the cryptographers is to implement universally
composable secure quantum communication schemes \cite{muller2009composability}.}

\textcolor{black}{From the discussion above it seems that the experimental
QKD is now a matured area. However, the same is not true for other
aspects of quantum cryptography, i.e., for the schemes beyond QKD
(e.g., schemes for two- and three-party secure direct quantum communication).
The secure direct quantum communication schemes exploit feasibility
of transmitting message in a secure manner without any requirement
of key generation/distribution, such as two-party one-way (quantum
secure direct communication (QSDC), deterministic secure quantum communication
(DSQC)), two-party two-way (quantum dialogue (QD)), three-party one-way
(controlled deterministic secure quantum communication (CDSQC)), three-party
two-way (controlled quantum dialogue (CQD)). Only a handful of experiments
have yet been performed. Specifically, QSDC has been realized with
single photons \cite{hu2016experimental} and entangled photons \cite{zhang2017quantum,zhu2017experimental}.
On top of that, quantum secret sharing has also been demonstrated
\cite{hai2013experimental,schmid2005experimental} and extended to
multiparty scenario as well \cite{smania2016experimental}. However,
our discussion is focused on secure direct quantum communication schemes
and quantum secret sharing is beyond the scope of the present work. }

\textcolor{black}{The above status of the experimental works have
motivated us to investigate possibilities of experimental realization
of quantum cryptographic schemes, such as QD \cite{pathak2013elements,nguyen2004quantum},
CQD \cite{thapliyal2015applications}, Kak\textquoteright s three-stage
scheme inspired direct communication scheme \cite{kak2006three,thapliyal2018kak},
CDSQC with entanglement swapping \cite{pathak2015efficient}, which
have not been experimentally realized so far. In the process, to design
optical schemes for the realization of these schemes, we have
perceived that the implementation
requires some modifications of the original schemes. Keeping this
point in mind, in the following sections of this work, we have modified
the original schemes which remain operationally equivalent to the
original scheme/without compromising with the security and have designed
optical circuits for the above mentioned quantum cryptograhic schemes,
which are based on single photon, two-qubit, and multi-qubit entangled
states (such as GHZ-like state, W state) using available optical elements,
like laser, beamsplitter (BS), polarizing beamsplitter (PBS), half-waveplate
(HWP).}

\textcolor{black}{The rest of the paper is organized as follows. In
Section \ref{sec:Quantum-cryptographic-protocols}, we have presented
the designs of optical circuits for various quantum cryptographic
tasks. Each circuit and the original protocol it implements are also
described in the corresponding section. Finally, the paper is concluded
in Section \ref{sec:Conclusions}.}

\section{\textcolor{black}{Quantum cryptographic protocols \label{sec:Quantum-cryptographic-protocols}}}

\textcolor{black}{In the previous section, we have already mentioned
that there exist unconditionally secure protocols for various quantum
communication tasks and a good number of experiments have been done.
However, until the recent past, experimental works on secure quantum
communication were restricted to the experimental realizations of
different protocols of QKD. Only recently (in 2016), a protocol of
QSDC was realized experimentally by Hu et al. \cite{hu2016experimental}.
Specifically, Hu et al., realized DL04 protocol \cite{deng2004secure}
using single-photon frequency coding. This pioneering work was a kind
of proof-of-principle table-top experiment. In this work, the requirement
of quantum memory was circumvented by delaying the photonic qubits
in the fiber coils. However, soon after Hu et al.'s pioneering work,
Zhang et al., \cite{zhang2017quantum} reported another experimental
realization of QSDC protocol through a table top experiment. Zhang
et al.'s experiment was fundamentally different from Hu et al.'s experiment
in two aspects- firstly Zhang et al., used entangled states and secondly
they used quantum memories based on atomic ensembles. Almost immediately
after the Zhang et al.'s experiment, in 2017, Zhu et al., reported
experimental realization of a QSDC scheme over a relatively longer
distance \cite{zhu2017experimental}. With these three experiments,
experimental quantum cryptography arrived at a stage beyond QKD, where
a set of schemes of two-party one-way secure direct communication
was experimentally realized using the available technologies. However,
there exist many multi-party schemes of secure direct quantum communication,
some of which are also two-way schemes. For example, any scheme for
QD would require two-way communication, whereas any scheme of controlled
quantum communication involves at least three parties (say, a scheme
for CQD). No such protocol has yet been realized experimentally. In
what follows, we will see that many of these protocols can be realized
experimentally using the existing technology. However, to do so, some
of the protocols would require some modifications, which are needed
for experimental realizations. Here, we would concentrate on such
suitably modified protocols, and the optical circuits which can be
used to experimentally realize those schemes. The optical designs
proposed here can also be used for experimental implementation of
some other DSQC schemes, for instance, MDI-QSDC scheme \cite{niu2018measurement}
using two-qubit entanglement and single photon source as well as Bell
measurement to accomplish required teleportation.}

\textcolor{black}{In the following }subs\textcolor{black}{ection,
we will briefly describe a protocol of CQD and how to implement that
using the existing optical technology. }\textcolor{black}{To do so,
we will be very precise and restrict ourselves from the detailed discussion
of the protocols or their security proof as those are available elsewhere
and those are not of the interest of the present work.}\textcolor{black}{{} Specifically,
we will briefly describe a protocol in a few steps which are essential.
We will also provide a clear schematic diagram of the optical setup
that can be used to realize the protocol, and will provide a step-wise
description of the working of the setup. The same strategy will be
followed in describing the other protocols, too.}

\subsection{\textcolor{black}{Controlled quantum dialogue}}

\textcolor{black}{To begin with, we may note that CQD is a three-party
scheme. In any CQD scheme, Alice and Bob want to exchange their secret
messages simultaneously to each other with the help of a third party
Charlie (controller). In what follows, we will first summarize a set
of CQD schemes of our interest \cite{thapliyal2017quantum,thapliyal2015applications}
and the bottleneck present in the implementation of the theoretical
schemes. After that we will explicitly show that it is possible to
design optical circuits to experimentally realize CQD with entangled
photons and single photon (in more than one way).}

\subsubsection{\textcolor{black}{Scheme T1: CQD with single photons}}

\textcolor{black}{Working of a CQD scheme based on single-qubit is
summarized in the following steps:}
\begin{description}
\item [{\textcolor{black}{Cs\_1:}}] \textcolor{black}{Charlie prepares
a random string of single qubits prepared in one of four states $|0\rangle,$
$|1\rangle,$ $|+\rangle$ and $|-\rangle$.}
\item [{\textcolor{black}{Cs\_2:}}] \textcolor{black}{Charlie sends the
string to Alice.}
\item [{\textcolor{black}{Cs\_3:}}] \textcolor{black}{After Alice confirms
that she has received the string, she randomly measures one-half of
the qubits either in $\left\{ |0\rangle,|1\rangle\right\} $ or $\left\{ |+\rangle,|-\rangle\right\} $
basis and announces her measurement basis and results with corresponding
position for checking the eavesdropping. Then Charlie compares the
Alice's measurement result with that of state preparation by using
classical communication (CC) in all the cases where they have chosen
the same basis}\footnote{\textcolor{black}{Throughout this} paper\textcolor{black}{, transmission
of qubits in all Schemes }\textbf{\textcolor{black}{Tj}}\textcolor{black}{{}
is performed along the same line after concatenation of a randomly
prepared string of decoy qubits followed by permutation of qubits
in the enlarged string. Subsequently, the error estimation on the
transmitted decoy qubits provides an upper bound of the errors introduced
during transmission on the remaining message qubits, which can be
solely attributed to the disturbance caused due to an eavesdropping
attempt for the sake of simplicity and attaining utmost security.
The choice of decoy qubits could be the single-qubit states used in
BB84 protocol or entangled states (see \cite{sharma2016verification}
for a detailed discussion).}}\textcolor{black}{.}
\item [{\textcolor{black}{Cs\_4:}}] \textcolor{black}{Alice encodes her
message on the one-half of the remaining qubits by using Pauli operations
$I$ or $iY$ to encode 0 or 1, respectively.}
\item [{\textcolor{black}{Cs\_5:}}] \textcolor{black}{Alice sends the message
encoded and (remaining) decoy qubits to Bob.}
\item [{\textcolor{black}{Cs\_6:}}] \textcolor{black}{Bob receives the
encoded photons (along with the remaining one-half to be used as decoy
qubits).}
\item [{\textcolor{black}{Cs\_7:}}] \textcolor{black}{To ensure the absence
of Eve, Alice discloses the positions of the decoy qubits and Charlie
announces corresponding choice of basis.}
\item [{\textcolor{black}{Cs\_8:}}] \textcolor{black}{Bob encodes his message
on the same qubits as was u}sed by Alice to encode her mes\textcolor{black}{sage.
Charlie announces the basis information for the message encoded qubits
when he wishes the task to be accomplished. After knowing the basis
of the initial state from Charlie, Bob measures the corresponding
qubits in that basis and announces the measurement outcomes. Using
the measurement outcomes both Alice and Bob can decode each other's
messages.}
\end{description}
\textcolor{black}{}
\begin{figure}
\begin{centering}
\textcolor{black}{\includegraphics[scale=0.43]{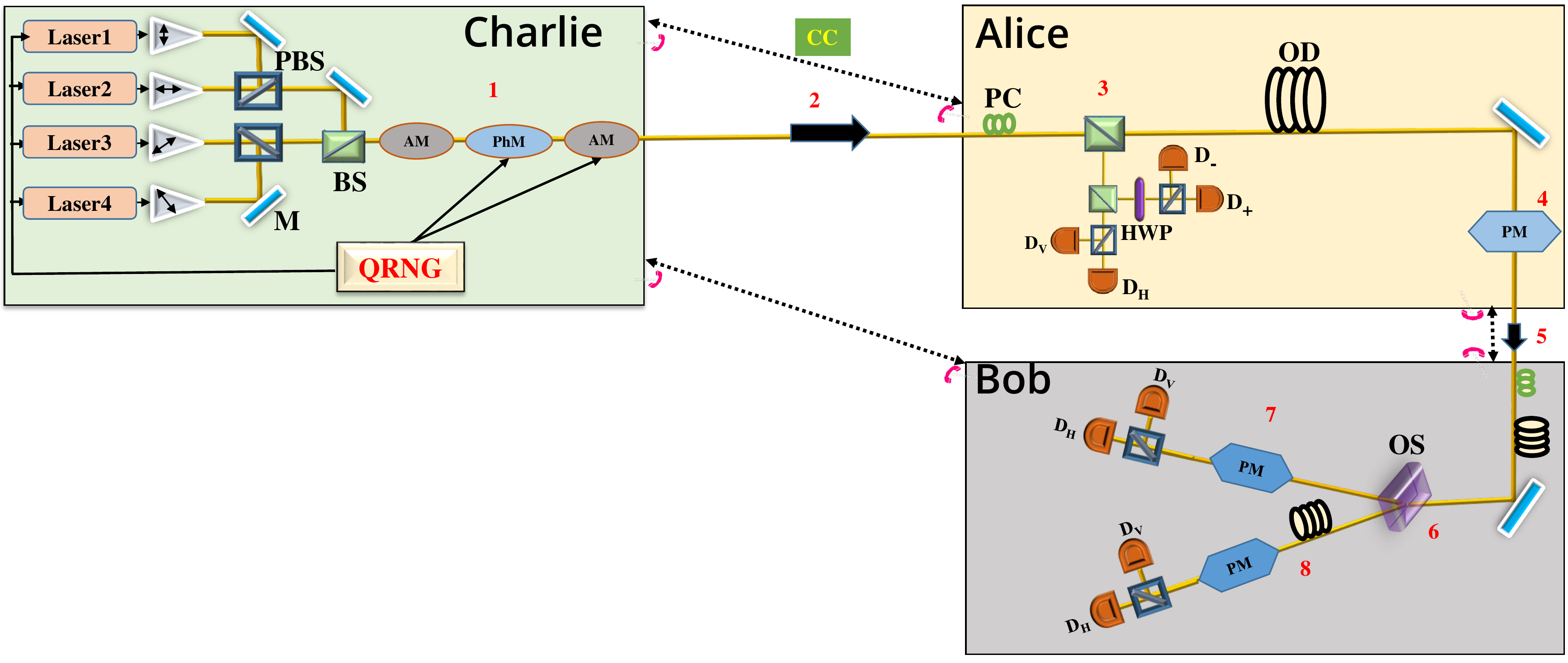}}
\par\end{centering}
\textcolor{black}{\caption{\label{fig:CQDS}Schematic diagram for CQD scheme based on polarization
qubit. Four lasers are used to prepare the polarization states of
photons. In the proposed optical design, BS stands for symmetric (50:50)
beam splitter, PBS is polarizing BS, M is mirror, AM is amplitude
modulator, PhM is phase modulator, QRNG is quantum random number generator,
PC is polarization controller, HWP is half wave plate, OD is optical
delay, PM is polarization modulator, OS is optical switch, $D_{i}$
represents the detector; whereas $D_{H}$ and $D_{V}$ correspond
to detectors used for measurements in horizontal and vertical basis
and similarly $D_{+}$ and $D_{-}$ correspond to measurements in
the diagonal basis; and CC is classical communication.}
}
\end{figure}
\textcolor{black}{The aforementioned CQD scheme has not yet been realized
experimentally due to unavailability of quantum memory, difficulty
in building on-demand single photon sources, permutation of particles,
and limitation in performing the scheme over scalable distance due
to the complexity of the task. However, there is a silver lining that
the CQD protocol using single photons can be realized using the existing
technology, such as using frequency encoding, and permutation of particles
can be circumvented at the cost of reduced qubit efficiency. To stress
on this point a schematic diagram of the optical setup that can be
used to realize the above protocol using polarization qubits is illustrated
in Fig. \ref{fig:CQDS} and in what follows, the same is elaborated
in a few steps.}

\subsubsection*{\textcolor{black}{Scheme E1: Optical design for CQD protocol using
single photons (polarization qubits)}}

The Scheme \textbf{T1} can be implemented in the following steps.
\begin{description}
\item [{\textcolor{black}{Cs-O\_1:}}] \textcolor{black}{Charlie uses four
lasers to generate the polarization state of the photons and tw}o
PBSs, \textcolor{black}{three mirrors, one symmetric (50:50) BS, two
amplitude modulators (AMs), one phase modulator (PhM), to generate
the random string of single photons in one of the four polarization
states $|H\rangle,$ $|V\rangle,$ $|+\rangle=\frac{|H\rangle+|V\rangle}{\sqrt{2}}$
and $|-\rangle=\frac{|H\rangle-|V\rangle}{\sqrt{2}}$. Whereas, the
first AM is used to generate the decoy photons, the second AM is used
to control the intensity of light, and PhM randomizes phase to generate
mixture of photon number states \cite{lo2014secure}. This is not
a unique method for the preparation of polarization qubit, which can
also be generated by heralding one outputs of the spontaneous parametric
down conversion (SPDC) outputs.}
\item [{\textcolor{black}{Cs-O\_2:}}] \textcolor{black}{Charlie sends the
string of single photons to Alice through optical fiber, open-air,
satellite, or under-water (in case of maritime cryptography). A polarization
controller is used to compensate for the changes in polarization in
case of transmission through an optical fiber.}
\item [{\textcolor{black}{Cs-O\_3:}}] \textcolor{black}{Alice receives
the string of photons and randomly selects one-half of the incoming
photons using a BS to check any eavesdropping attempt. She randomly
measures the reflected photons either in $\left\{ |H\rangle,\,|V\rangle\right\} $
or in $\left\{ |+\rangle,|-\rangle\right\} $ basis (again using a
BS) and announces her measurement basis and results with corresponding
position for checking the eavesdropping. Then Charlie compares Alice's
measurement result with that of state preparation. While the eavesdropping
checking between Charlie and Alice, she uses an optical delay (serving
as a quantum memory) for the rest of the photons.}\\
\textcolor{black}{Suppose Charlie prepares the photon from Laser1,
i.e., in state $|V\rangle$ then detector $D_{V}$ is expected to
click in the ideal case, but if the detector $D_{H}$ clicks, then
it will be registered as bit error. However, if the detector $D_{+}$
or $D_{-}$ clicks then this case will be discarded.}
\item [{\textcolor{black}{Cs-O\_4:}}] \textcolor{black}{Alice encodes her
message on one-half of the transmitted photons by using a polarization
modulator (PM) or a set of a half-waveplate sandwiched between two
quarter-wave plates.}
\item [{\textcolor{black}{Cs-O\_5:}}] \textcolor{black}{Alice sends encoded
and decoy photons to Bob.}
\item [{\textcolor{black}{Cs-O\_6:}}] \textcolor{black}{Bob receives the
encoded photons along with the decoy photons and keeps the received
photons in an optical delay. Subsequently, Alice discloses the positions
of the decoy qubits, Charlie discloses corresponding basis, and Bob
passes the string of photons through an optical switch which sends
the encoded photons and decoy photons on different paths.}
\item [{\textcolor{black}{Cs-O\_7:}}] \textcolor{black}{To ensure the absence
of Eve, Bob chooses the basis of the decoy qubits by using a PM and
measures them to compute the error rate. They proceed if the errors
are below threshold.}
\item [{\textcolor{black}{Cs-O\_8:}}] \textcolor{black}{Bob encodes his
message using a PM on the same photons used by Alice to encode his
message. Subsequently, Charlie reveals the basis information of corresponding
state preparation. After knowing the basis choice of the initial state
from Charlie, he measures the message encoded photons using two single
photon detectors and a PM to choose the basis of the states to be
measured and announces his result. In fact, Bob can perform the same
task using only one PM if he delays his encoding till Charlie reveals
the basis information. From the measurement outcomes both Alice and
Bob will be able to decode each other's messages.}
\end{description}

\subsubsection{\textcolor{black}{Scheme T2: Kak\textquoteright s three-stage scheme
inspired five-stage scheme of CQD with single photons}}

\begin{figure}
\begin{centering}
\textcolor{black}{\includegraphics[scale=0.47]{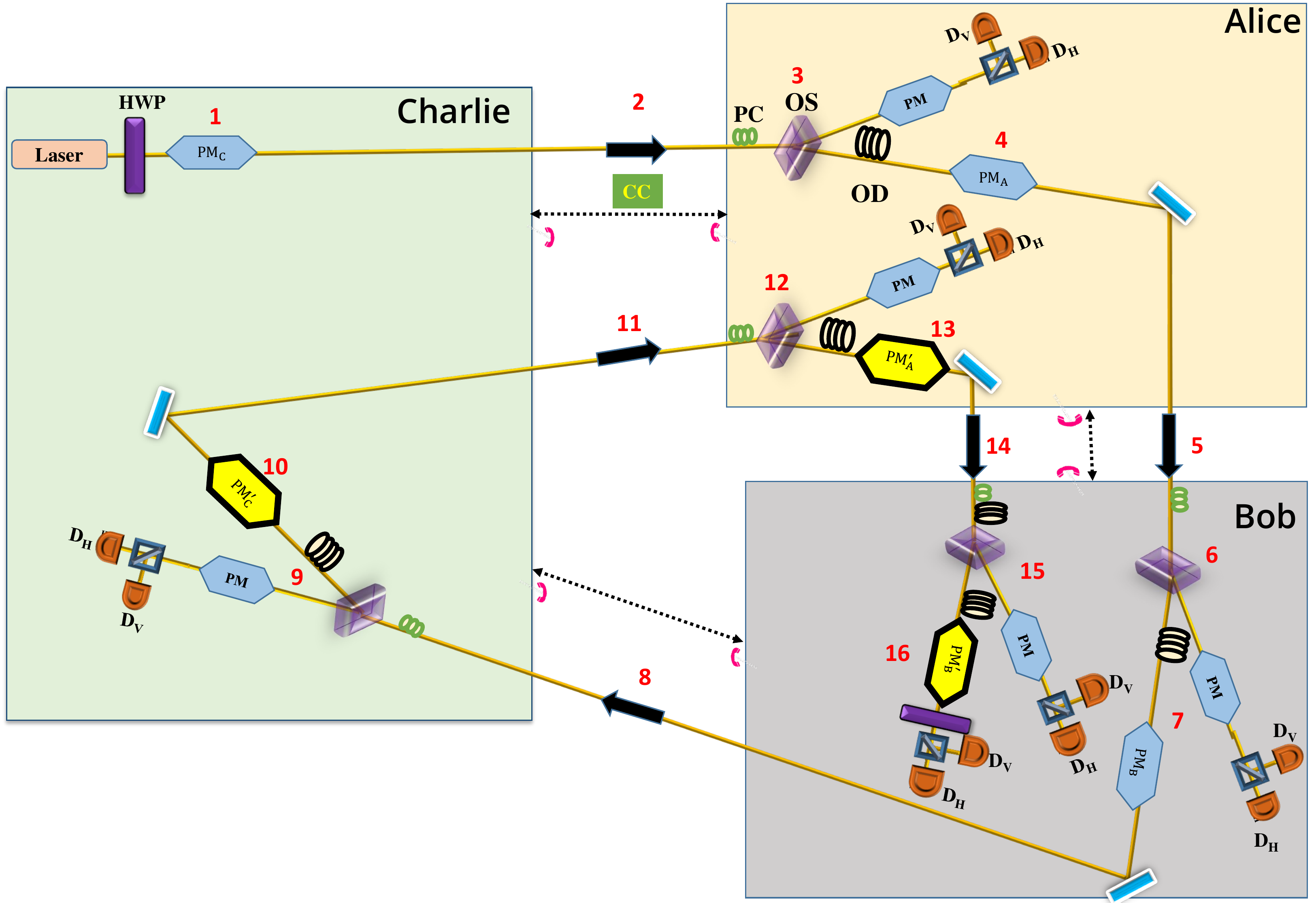}}
\par\end{centering}
\textcolor{black}{\caption{\label{fig:KAK_MODIFIED}A proposed optical implementation using polarization
qubit of the five-step CQD scheme inspired from the three-stage scheme.
Laser is used to prepare the polarization encoded qubits. In all the
lab's, ${\rm PM_{A,B,C}}$ is polarization modulator used to implement
a rotation operator, and ${\rm PM_{A,B,C}^{\prime}}$ performs corresponding
inverse rotation operator.}
}
\end{figure}

\textcolor{black}{A three-stage QKD scheme proposed in the past \cite{kak2006three}
was recently shown to be able to perform secure direct quantum communication
\cite{thapliyal2018kak}. Here, we propose a three-stage scheme inspired
CQD protocol, which can be viewed as a five-stage protocol of CQD.
The protocol is summarized in the following steps:}
\begin{description}
\item [{\textcolor{black}{CK\_1:}}] \textcolor{black}{Charlie prepares
a string of single qubits in the computational basis. Subsequently,
he applies random unitary operators on each qubit and keeps the corresponding
information with himself.}
\item [{\textcolor{black}{CK\_2:}}] \textcolor{black}{Same as}\textbf{\textcolor{black}{{}
Cs\_2}}\textcolor{black}{.}
\item [{\textcolor{black}{CK\_3:}}] \textcolor{black}{After Alice confirms
that she has received the qubits she measures one-half of the received
qubits to check eavesdropping chosen by Charlie, who also disclose
corresponding rotation operator and the initial state.}
\item [{\textcolor{black}{CK\_4:}}] \textcolor{black}{Alice applies a random
rotation operator on all the qubits.}
\item [{\textcolor{black}{CK\_5:}}] \textcolor{black}{Same as}\textbf{\textcolor{black}{{}
Cs\_5}}\textcolor{black}{.}
\item [{\textcolor{black}{CK\_6:}}] \textcolor{black}{Same as}\textbf{\textcolor{black}{{}
CK\_3}}\textcolor{black}{, here Bob measures one-half of the received
qubits with the help of information of rotation operators by Charlie
and Alice as well as the initial state revealed by Charlie.}
\item [{\textcolor{black}{CK\_7:}}] \textcolor{black}{Same as}\textbf{\textcolor{black}{{}
CK\_4}}\textcolor{black}{, but Bob applies his rotation operator.}
\item [{\textcolor{black}{CK\_8:}}] \textcolor{black}{Bob sends all the
qubits to Charlie.}
\item [{\textcolor{black}{CK\_9:}}] \textcolor{black}{After Charlie confirms
that he has received the qubits he measures one-half of the received
qubits to check eavesdropping with the help of Alice's, Bob's, and
his own rotation operators.}
\item [{\textcolor{black}{CK\_10:}}] \textcolor{black}{Charlie applies
an inverse of his rotation operator applied in }\textbf{\textcolor{black}{CK\_1}}\textcolor{black}{.}
\item [{\textcolor{black}{CK\_11:}}] \textcolor{black}{Same as}\textbf{\textcolor{black}{{}
Cs\_2}}\textcolor{black}{.}
\item [{\textcolor{black}{CK\_12:}}] \textcolor{black}{Same as}\textbf{\textcolor{black}{{}
CK\_3}}\textcolor{black}{, but here Alice requires information of
the rotation operator from Bob.}
\item [{\textcolor{black}{CK\_13:}}] \textcolor{black}{Alice applies inverse
of the rotation operator applied in }\textbf{\textcolor{black}{CK\_4}}\textcolor{black}{.
Subsequently, she also encodes her message on one-half of the remaining
qubits using Pauli operations $I$ or $X$ to send 0 or 1, respectively.}
\item [{\textcolor{black}{CK\_14:}}] \textcolor{black}{Same as}\textbf{\textcolor{black}{{}
Cs\_5}}\textcolor{black}{.}
\item [{\textcolor{black}{CK\_15:}}] \textcolor{black}{Same as}\textbf{\textcolor{black}{{}
CK\_6}}\textcolor{black}{, but Bob needs only his rotation operator,
while positions of the decoy qubits are disclosed by Alice and initial
state by Charlie.}
\item [{\textcolor{black}{CK\_16:}}] \textcolor{black}{Bob applies inverse
of his rotation operator and encodes his message on all the qubits.
He subsequently measures the transformed qubits in the computational
basis and announces the measurement outcomes. Finally, Charlie reveals
the initial states when he wishes them to accomplish the task. With
the help of the initial and final states both Alice and Bob can decode
each other's messages.}
\end{description}
\textcolor{black}{To complete two rounds, first for locking and second
for u}nlocking, between three parties the qubits should travel five
times through the lossy transmission channel which sets limitations
on the experimental implementation of the present scheme. However,
to remain consistent with the theme of the present work, where we
aim to discuss the possibilities of reduction of complex\textcolor{black}{{}
quantum cryptographic tasks to obtain the solutions of simpler secure
communication tasks, we now discuss the optically implementable scheme.
In principle, the protocol described here can also be realized using
available optical elements and a schematic diagram for that is shown
in Figure \ref{fig:KAK_MODIFIED}, and the same is described below
in a few steps.}

\subsubsection*{\textcolor{black}{Scheme E2: Optical design for five-stage scheme
using single photons (polarization qubits)}}

Optical implementation of Scheme \textbf{T2} works as follows.
\begin{description}
\item [{\textcolor{black}{CK-O\_1:}}] \textcolor{black}{Charlie prepares
a random string of horizontal and vertical polarized single photons
and uses ${\rm PM_{C}}$ to rotate the polarization of these photons}\textcolor{black}{{} randomly.}
\item [{\textcolor{black}{CK-O\_2:}}] \textcolor{black}{Same as}\textbf{\textcolor{black}{{}
Cs-O\_2}}\textcolor{black}{.}
\item [{\textcolor{black}{CK-O\_3:}}] \textcolor{black}{Same as}\textbf{\textcolor{black}{{}
Cs-O\_3}}\textcolor{black}{, but here Charlie informs the decoy qubits
by revealing the positions and corresponding random unitary operation
using which Alice measures the photons in the computational basis
with the help of optical switch, PM and single photon detectors. They
proceed only if fewer than the threshold bit-flip errors are observed. }
\begin{figure}
\begin{centering}
\textcolor{black}{\includegraphics[scale=0.43]{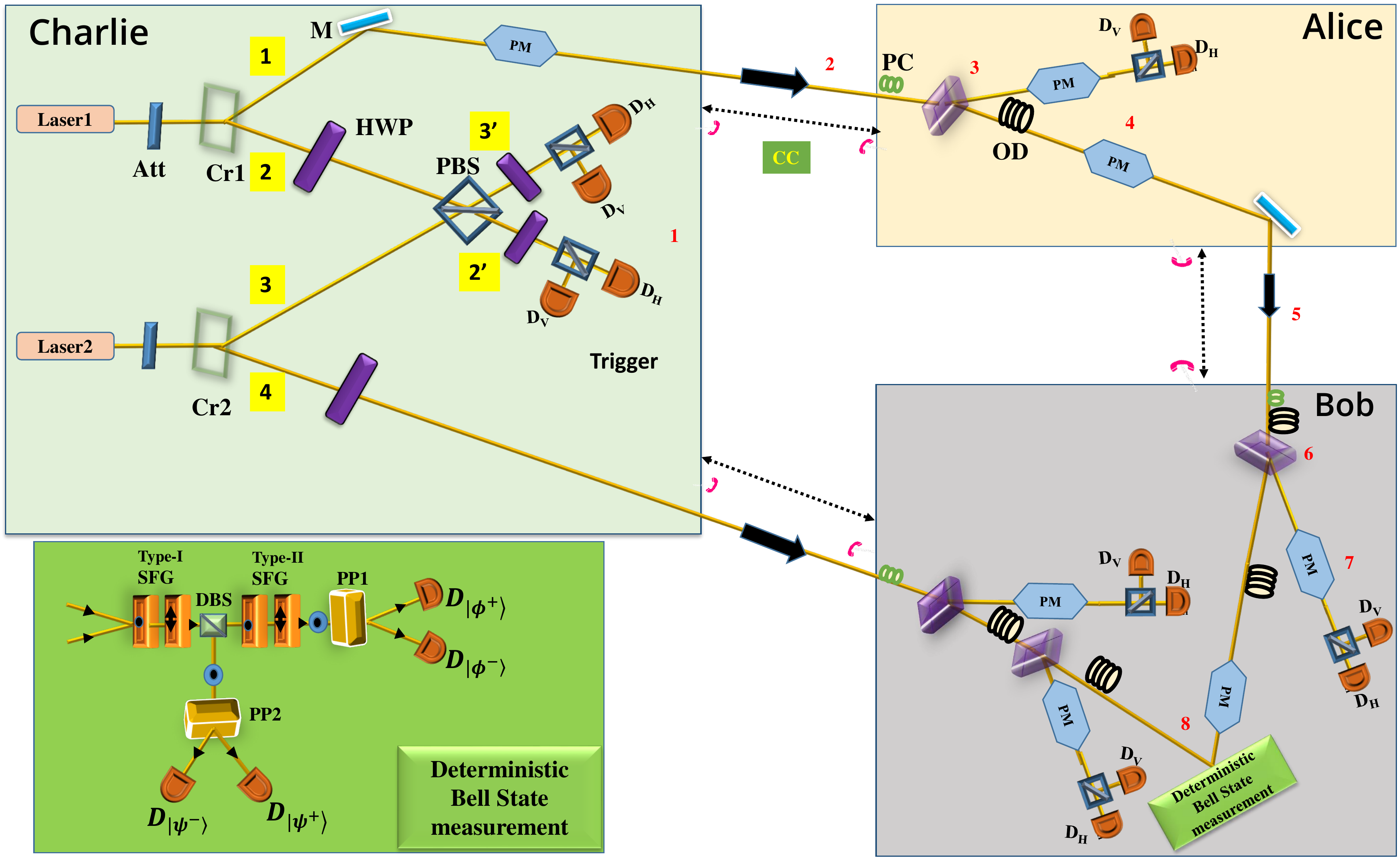}}
\par\end{centering}
\textcolor{black}{\caption{\label{fig:CQD_entangled}Optical design of the CQD scheme with entangled
photons and a complete Bell state measurement. In Charlie's lab Cr1
and Cr2 are two nonlinear crystals which are used to generate the
entangled photons. Attenuator (Att) is used to control the intensity
of light so that a single Bell pair can be generated. This can also
be controlled by changing the pump power. Sum frequency generation
(SFG) type-I and type-II are nonlinear interactions, which are used
to perform the Bell state measurement. DBS is dichroic BS and PP1
and PP2 are $45^{0}$ projector. BSM is shown in the box.}
}
\end{figure}
\item [{\textcolor{black}{CK-O\_4:}}] \textcolor{black}{Alice applies a
random rotation on the rest of the qubits using ${\rm PM_{A}}$.}
\item [{\textcolor{black}{CK-O\_5:}}] \textcolor{black}{Same as}\textbf{\textcolor{black}{{}
Cs-O\_5}}\textcolor{black}{.}
\item [{\textcolor{black}{CK-O\_6:}}] \textcolor{black}{Same as}\textbf{\textcolor{black}{{}
CK-O\_3}}\textcolor{black}{, here Alice and Bob perform eavesdropping
checking with the help of Charlie.}
\item [{\textcolor{black}{CK-O\_7:}}] \textcolor{black}{Same as}\textbf{\textcolor{black}{{}
CK-O\_4}}\textcolor{black}{, but Bob applies the random operation
using ${\rm PM_{B}}$.}
\item [{\textcolor{black}{CK-O\_8:}}] \textcolor{black}{Bob sends all the
qubits to Charlie.}
\item [{\textcolor{black}{CK-O\_9:}}] \textcolor{black}{Same as}\textbf{\textcolor{black}{{}
CK-O\_3}}\textcolor{black}{, here Charlie needs assistance of both
Alice and Bob to perform eavesdropping checking.}
\item [{\textcolor{black}{CK-O\_10:}}] \textcolor{black}{Charlie applies
the inverse of his rotation operator}\textbf{\textcolor{black}{{} }}\textcolor{black}{${\rm PM_{C}^{\prime}}$.}
\item [{\textcolor{black}{CK-O\_11-12:}}] \textcolor{black}{Same as}\textbf{\textcolor{black}{{}
CK-O\_2-3}}\textcolor{black}{.}
\item [{\textcolor{black}{CK-O\_13:}}] \textcolor{black}{Alice applies
an inverse operation of her rotation operator}\textbf{\textcolor{black}{{}
}}\textcolor{black}{using ${\rm PM_{A}^{\prime}}$. She also encodes
her message on one-half of the qubits in this step (by using ${\rm PM_{A}^{\prime}}$
only).}
\item [{\textcolor{black}{CK-O\_14-15:}}] \textcolor{black}{Same as}\textbf{\textcolor{black}{{}
CK-O\_5-6}}\textcolor{black}{, but positions of decoy qubits are disclosed
by Alice.}
\item [{\textcolor{black}{CK-O\_16:}}] \textcolor{black}{Bob applies his
inverse rotation operator ${\rm PM_{B}^{\prime}}$ and encodes his
message. Then he measures the polarization of the transformed qubits
and announces the result. When Charlie wishes them to complete the
task, he reveals the initial choice of polarization of his qubit.}
\end{description}
\textcolor{black}{So far, we have discussed schemes of CQD using single
photons. Extending the idea, in what follows, we will discuss the
CQD scheme using entangled states and subsequently discuss the optical
implementation of such scheme.}

\textcolor{black}{}

\subsubsection{\textcolor{black}{Scheme T3: CQD protocol with entangled photons}}

\textcolor{black}{CQD scheme based on entangled qubits is summarized
in the following steps:}
\begin{description}
\item [{\textcolor{black}{Ce\_1:}}] \textcolor{black}{Charlie prepares
a string of three qubit GHZ-like states $|\psi\rangle_{123}=|\psi^{+}0\rangle+|\phi^{+}1\rangle$
with $\left\{ |\psi^{\pm}\rangle=\frac{|00\rangle\pm|11\rangle}{\sqrt{2}},|\phi^{\pm}\rangle=\frac{|01\rangle\pm|10\rangle}{\sqrt{2}}\right\} $.
He uses first and second qubit as a travel qubit and third qubit as
a home qubit.}
\item [{\textcolor{black}{Ce\_2:}}] \textcolor{black}{Charlie keeps all
the third qubits and sends the strings of first and second qubits
to Alice and Bob, respectively after inserting some decoy qubits in
each string.}
\item [{\textcolor{black}{Ce\_3:}}] \textcolor{black}{Same as }\textbf{\textcolor{black}{Cs\_3}}\textcolor{black}{,
but here both Alice and Bob perform eavesdropping checking on the
strings received from Charlie.}
\item [{\textcolor{black}{Ce\_4:}}] \textcolor{black}{Same as }\textbf{\textcolor{black}{Cs\_4}}\textcolor{black}{,
but Alice can use dense coding and encode 2 bits of message using
all four Pauli operations.}
\item [{\textcolor{black}{Ce\_5:}}] \textcolor{black}{Same as }\textbf{\textcolor{black}{Cs\_5}}\textcolor{black}{,
but here Alice randomly inserts freshly prepared decoy qubits before
sending message encoded qubits to Bob.}
\item [{\textcolor{black}{Ce\_6:}}] \textcolor{black}{Same as }\textbf{\textcolor{black}{Cs\_6}}\textcolor{black}{.}
\item [{\textcolor{black}{Ce\_7:}}] \textcolor{black}{Same as }\textbf{\textcolor{black}{Cs\_7}}\textcolor{black}{,
but Charlie need not disclose anything. }
\item [{\textcolor{black}{Ce\_8:}}] \textcolor{black}{Bob encodes his message
on the same qubits which were used by Alice to encode }her \textcolor{black}{message.
Subsequently, Bob performs the Bell measurement on the messages encoded
string and the string received from Charlie. Then he announces the
measurement results. Now, Alice and Bob decode the message with the
help of Charlie's measurement results of the third qubits.}
\end{description}
\textcolor{black}{The CQD protocol using entangled photons described
above can be realized using optical elements. A schematic diagram
for that is shown in Figure \ref{fig:CQD_entangled}, which is described
briefly below in a few steps.}

\subsubsection*{\textcolor{black}{Scheme E3: Optical design for CQD protocol using
entangled photons }(polarization qubits)}

Scheme \textbf{T3} can be implemented as
\begin{description}
\item [{\textcolor{black}{Ce-O\_1:}}] \textcolor{black}{Charlie uses two
lasers and two non-linear crystals (Cr1 and Cr2) to generate the two
pairs of Bell states with the help of SPDC process.
\[
\begin{array}{ccc}
|\psi\rangle_{1} & = & \left(\frac{|HH\rangle+|VV\rangle}{\sqrt{2}}\right)_{12}\otimes\left(\frac{|HH\rangle+|VV\rangle}{\sqrt{2}}\right)_{34},\end{array}
\]
}
\end{description}
\textcolor{black}{where $H$ and $V$ represent the horizontal and
vertical polarizations, respectively.}

\textcolor{black}{Now, $2$nd and $4$th photons pass through the
HWP $\left(2\theta=45^{0}\right)$ and the state becomes}
\textcolor{black}{
\begin{equation}
\begin{array}{lcl}
|\psi\rangle_{2} & = & \left(\frac{|H+\rangle+|V-\rangle}{\sqrt{2}}\right)_{12}\otimes\left(\frac{|H+\rangle+|V-\rangle}{\sqrt{2}}\right)_{34}\\
 & = & \frac{1}{2}\left(|H+H+\rangle+|H+V-\rangle+|V-H+\rangle+|V-V-\rangle\right)_{1234}.
\end{array}\label{eq:-4}
\end{equation}}
Subsequently, the $2$nd and $4$th photons pass through the two input
ports of a PBS (which transmits the horizontal photon and reflects
the vertical photon). The postselected state after passing through
PBS such that only one photon will be in each output path can be written
after renormalization as
\textcolor{black}{
\[
\begin{array}{ccc}
|\psi\rangle_{3} & = & \frac{1}{2}\left(|HHH+\rangle+|HVV-\rangle+|VHH+\rangle-|VVV-\rangle\right)_{12^{'}3^{'}4}\end{array}.
\]
Then photon $2^{'}$ passes through a HWP $\left(2\theta=45^{0}\right)$
and state transforms as}

\textcolor{black}{
\begin{equation}
\begin{array}{lcl}
|\psi\rangle_{4} & = & \frac{1}{2}\left(|+\rangle_{2^{'}}\left(|+H+\rangle+|-V-\rangle\right)_{13^{'}4}+|-\rangle_{2^{'}}\left(|+H+\rangle-|-V-\rangle\right)_{13^{'}4}\right)\\
 & = & \frac{1}{2}\left(|+\rangle_{2^{'}}\left(|+\rangle_{3^{'}}|\psi^{+}\rangle_{14}+|-\rangle|\phi^{+}\rangle_{14}\right)+|-\rangle_{2^{'}}\left(|+\rangle_{3^{'}}|\phi^{+}\rangle_{14}+|-\rangle|\psi^{+}\rangle_{14}\right)\right)\\
 & = & \frac{1}{2}\left(|+\rangle_{2^{'}}\left(|\Phi_{1}\rangle\right)+|-\rangle_{2^{'}}\left(|\Phi_{2}\rangle\right)\right).
\end{array}\label{eq:-2-1}
\end{equation}
}

\textcolor{black}{From the obtained state one can clearly see that
if Charlie measures qubit $2^{'}$ and announces the measurement outcome,
depending upon which all the parties can decide which channel they
are sharing. Otherwise, if Charlie measures $|+\rangle_{2^{'}}$,
then he will get $|+\rangle_{3^{'}}|\psi^{+}\rangle_{14}+|-\rangle|\phi^{+}\rangle_{14}$,
need not apply any gate, i.e., identity, if he measures $|-\rangle_{2^{'}}$,
then he will get $|+\rangle_{3^{'}}|\phi^{+}\rangle_{14}+|-\rangle|\psi^{+}\rangle_{14}$,
he applies NOT gate on $1$ or $4$, then the state will be $|+\rangle_{3^{'}}|\psi^{+}\rangle_{14}+|-\rangle|\phi^{+}\rangle_{14}.$
Also note that $|\Phi_{i}\rangle$ are the unitary equivalent of the
state prepared in }\textbf{\textcolor{black}{Ce\_1}}\textcolor{black}{.
The experimental preparation of three-qubit states using this approach
can be found in Ref. \cite{zhang2006experimental}.}
\begin{description}
\item [{\textcolor{black}{Ce-O\_2:}}] \textcolor{black}{Charlie sends corresponding
photons 1 and 4 to Alice and Bob, respectively.}
\item [{\textcolor{black}{Ce-O\_3:}}] \textcolor{black}{Both Alice and
Bob receive the photons and choose the same set of photons using an
optical switch to check their correlations with Charlie to check the
eavesdropping. Bob keeps her photons in an optical delay.  Here, it
is worth mentioning that Alice and Bob can also use BS (as in }\textbf{\textcolor{black}{Cs-O\_3}}\textcolor{black}{)
for this task, but that will reduce qubit efficiency as the cases
where one of them has measured entangled state, but not other will
be discarded.}
\item [{\textcolor{black}{Ce-O\_4:}}] \textcolor{black}{Same as }\textbf{\textcolor{black}{Cs-O\_4}}\textcolor{black}{,
but Alice uses dense coding as well.}
\item [{\textcolor{black}{Ce-O\_5-6:}}] \textcolor{black}{Same as }\textbf{\textcolor{black}{Cs-O\_5}}\textcolor{black}{-}\textbf{\textcolor{black}{6}}\textcolor{black}{.}
\item [{\textcolor{black}{Ce-O\_7:}}] \textcolor{black}{Same as }\textbf{\textcolor{black}{Cs-O\_7}}\textcolor{black}{,
but Charlie has to reveal the measurement outcome for the corresponding
decoy qubits $3^{'}$ in diagonal basis.}
\item [{\textcolor{black}{Ce-O\_8:}}] \textcolor{black}{Same as }\textbf{\textcolor{black}{Cs-O\_8}}\textcolor{black}{,
but Bob encodes his 2 bits of message on each photon used by Alice
to encode her message by using PM. After that, Bob performs Bell measurement
\cite{kim2001quantum} and announces the measurement outcomes. Similarly,
Charlie measures his qubit in the diagonal basis and announces the
measurement results when he wishes Alice and Bob to decode the messages.
There are schemes for probabilistic Bell measurement, but is not desirable
in the implementation of direct communication schemes as it is prone
to loss in encoded message. The drawback of deterministic Bell state
measurement is small efficiency due to involvement of nonlinear optics
in its implementation.}
\end{description}
In principle, Charlie can measure qubit \textcolor{black}{$3^{'}$
in }\textbf{\textcolor{black}{Ce-O\_1 }}\textcolor{black}{and keep
measurement result until }\textbf{\textcolor{black}{Ce-O\_7-8}}\textcolor{black}{.}

\subsection{\textcolor{black}{Controlled direct secure quantum communication}}

\textcolor{black}{There is three-party one-way another controlled
communication scheme where QSDC/DSQC from Alice to Bob is supervised
by a controller. Specifically, Alice can directly transmits the message
in a secure manner to Bob with the help of controller. QSDC (DSQC)
scheme require no (an) additional CC by receiver for decoding the
message. However, their controlled counterparts fall in CDSQC because
Charlie has to disclose his information in the end.}

\subsubsection{\textcolor{black}{CDSQC with single photons}}

\textcolor{black}{The CQD schemes }\textbf{\textcolor{black}{T1}}\textcolor{black}{,
}\textbf{\textcolor{black}{T2}}\textcolor{black}{{} (}\textbf{\textcolor{black}{E1}}\textcolor{black}{,
}\textbf{\textcolor{black}{E2}}\textcolor{black}{) discussed in the
previous section can be reduced to CDSQC schemes if Bob does not encode
his message in the last step. Additionally, he need not announce the
measurement outcome as he is not sending message to Alice in this
task.}

\subsubsection{\textcolor{black}{CDSQC with entangled photons}}

\textcolor{black}{The entangled states based CQD scheme }\textbf{\textcolor{black}{T3}}\textcolor{black}{{}
(}\textbf{\textcolor{black}{E3}}\textcolor{black}{) can also be reduced
analogous to single photon based scheme to obtain a CDSQC scheme.
Here, we have presented another entangled state based CDSQC \ref{fig:CDSQC}
with entanglem}ent swapping, where message encoded qubits are not
accessible to Eve as those qubits do not travel through the channel.

\subsubsection{\textcolor{black}{Scheme T4: CDSQC with entanglement swapping}}

\begin{figure}
\begin{centering}
\includegraphics[scale=0.4]{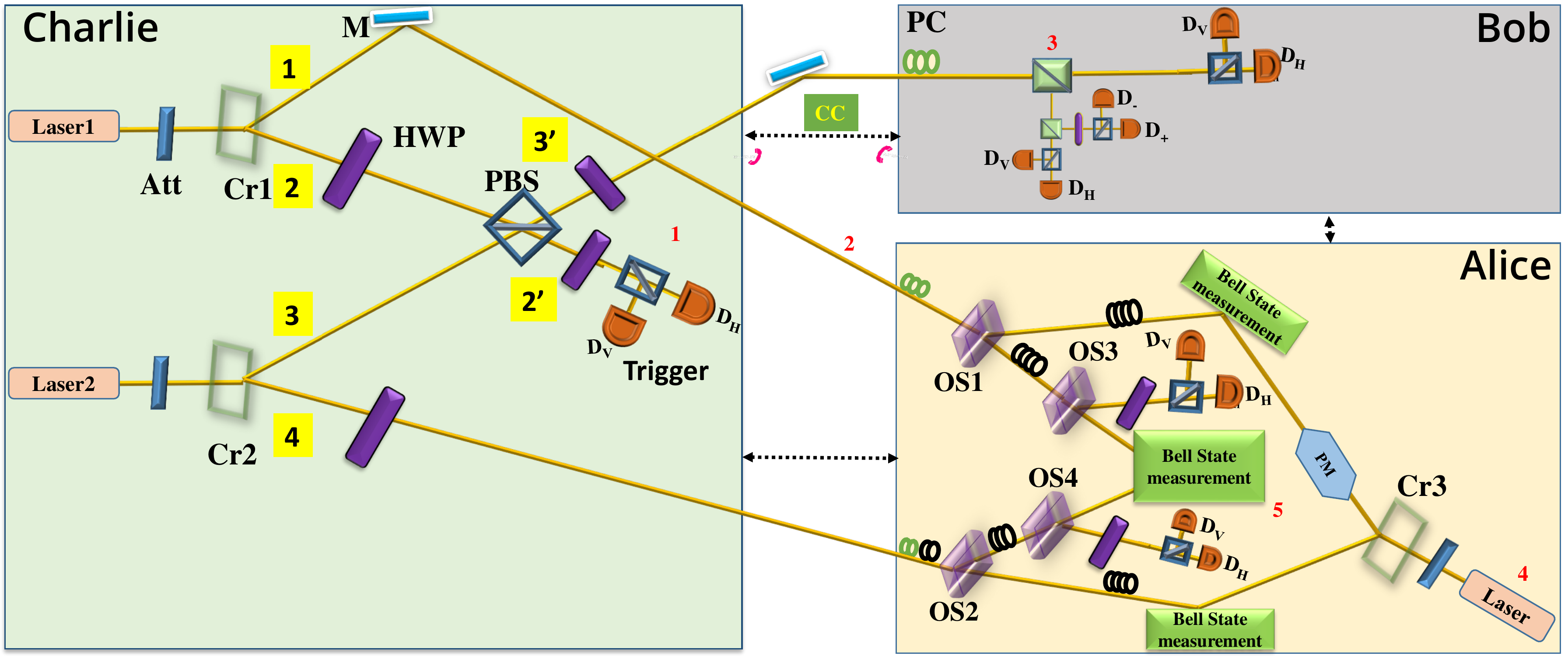}
\par\end{centering}
\caption{\label{fig:CDSQC}Schematic optical design of controlled direct secure
quantum communication with entanglement swapping. Nonlinear crystal
Cr3 is used by Alice for generation of Bell state.}
\end{figure}

\textcolor{black}{CDSQC with entanglement swapping is summarized in
the following steps:}
\begin{description}
\item [{\textcolor{black}{D\_1:}}] \textcolor{black}{Charlie prepares a
four qubit entangled state 
\begin{equation}
|\psi\rangle=\frac{1}{2}\left\{ \left(|\psi^{+}\rangle_{12}|0\rangle_{3}+|\phi^{+}\rangle_{12}|1\rangle_{3}\right)|0\rangle_{4}+\left(|\phi^{+}\rangle_{12}|0\rangle_{3}+|\psi^{+}\rangle_{12}|1\rangle_{3}\right)|1\rangle_{4}\right\} ,\label{eq:4-qubit state}
\end{equation}
}
\end{description}
\textcolor{black}{where qubit $4$ corresponds to Charlie, qubits
$1$ and $2$ for Alice and $3$ for Bob.}
\begin{description}
\item [{\textcolor{black}{D\_2:}}] \textcolor{black}{Same as }\textbf{\textcolor{black}{Ce\_2
}}\textcolor{black}{but Charlie keeps the fourth qubit, sends the
first and second qubits to Alice and third qubit to Bob.}
\item [{\textcolor{black}{D\_3:}}] \textcolor{black}{Same as }\textbf{\textcolor{black}{Ce\_3}}\textcolor{black}{.}
\item [{\textcolor{black}{D\_4:}}] \textcolor{black}{Alice prepares $|\psi^{+}\rangle_{A_{1}A_{2}}$
to encode her secret information. Specifically, she encodes 1 (0)
by a}pplying a $X$ ($I)$ gate on one of the qubits of the Bell state.
Thus, the combined state becomes\textcolor{black}{
\begin{equation}
\begin{array}{lcl}
|\psi^{\prime}\rangle & = & \frac{1}{2\sqrt{2}}\left(|\psi^{+}\rangle_{A_{1}A_{2}}\left(|\psi^{+}\rangle_{12}|0\rangle_{3}+|\phi^{+}\rangle_{12}|1\rangle_{3}\right)|0\rangle_{4}\right.\\
 & + & \left.|\psi^{+}\rangle_{A_{1}A_{2}}\left(|\phi^{+}\rangle_{12}|0\rangle_{3}+|\psi^{+}\rangle_{12}|1\rangle_{3}\right)|1\rangle_{4}\right)
\end{array}.\label{eq:-5}
\end{equation}
}
\item [{\textcolor{black}{D\_5:}}] \textcolor{black}{Alice measures qubits
$A_{1}$ and 1 as well as $A_{2}$ and 2 in the Bell basis, while
Bob and Charlie can measure their qubits in the computational basis.
Subsequently, Alice and Charlie announce their measurement outcomes,
which should reveal Alice's message to Bob.}\\
\textcolor{black}{To illustrate this point we can write the state
before Alice's and Bob's measurements, while Charlie's measurement
re}sult is $|0\rangle$ and Alice encodes $0$, as \textcolor{black}{
\begin{equation}
\begin{array}{lcl}
|\psi^{\prime}\rangle & = & \frac{1}{2\sqrt{2}}\left(\left\{ |\psi^{+}\rangle_{A_{1}1}|\psi^{+}\rangle_{A_{2}2}+|\phi^{+}\rangle_{A_{1}1}|\phi^{+}\rangle_{A_{2}2}+|\phi^{-}\rangle_{A_{1}1}|\phi^{-}\rangle_{A_{2}2}+|\psi^{-}\rangle_{A_{1}1}|\psi^{-}\rangle_{A_{2}2}\right\} |0\rangle_{3}\right.\\
 & + & \left.\left\{ |\psi^{+}\rangle_{A_{1}1}|\phi^{+}\rangle_{A_{2}2}+|\psi^{-}\rangle_{A_{1}1}|\phi^{-}\rangle_{A_{2}2}+|\phi^{+}\rangle_{A_{1}1}|\psi^{+}\rangle_{A_{2}2}+|\phi^{-}\rangle_{A_{1}1}|\psi^{-}\rangle_{A_{2}2}\right\} |1\rangle_{3}\right).
\end{array}\label{eq:msg-1}
\end{equation}
}Similarly, if \textcolor{black}{Charlie's measurement result is $|1\rangle$
and Alice encodes $1$, the reduced state will be
\begin{equation}
\begin{array}{lcl}
|\psi^{\prime}\rangle & = & \frac{1}{2\sqrt{2}}\left(\left\{ |\psi^{+}\rangle_{A_{1}1}|\phi^{+}\rangle_{A_{2}2}+|\phi^{+}\rangle_{A_{1}1}|\psi^{+}\rangle_{A_{2}2}-|\phi^{-}\rangle_{A_{1}1}|\psi^{-}\rangle_{A_{2}2}-|\psi^{-}\rangle_{A_{1}1}|\phi^{-}\rangle_{A_{2}2}\right\} |1\rangle_{3}\right.\\
 & + & \left.\left\{ |\psi^{+}\rangle_{A_{1}1}|\psi^{+}\rangle_{A_{2}2}-|\psi^{-}\rangle_{A_{1}1}|\psi^{-}\rangle_{A_{2}2}+|\phi^{+}\rangle_{A_{1}1}|\phi^{+}\rangle_{A_{2}2}-|\phi^{-}\rangle_{A_{1}1}|\phi^{-}\rangle_{A_{2}2}\right\} |0\rangle_{3}\right).
\end{array}\label{eq:19-1}
\end{equation}
If Charlie's measurement result is $|0\rangle$ and Alice encodes
$1$, the reduced state is
\begin{equation}
\begin{array}{lcl}
|\psi^{\prime}\rangle & = & \frac{1}{2\sqrt{2}}\left(\left\{ |\psi^{+}\rangle_{A_{1}1}|\psi^{+}\rangle_{A_{2}2}-|\psi^{-}\rangle_{A_{1}1}|\psi^{-}\rangle_{A_{2}2}+|\phi^{+}\rangle_{A_{1}1}|\phi^{+}\rangle_{A_{2}2}-|\phi^{-}\rangle_{A_{1}1}|\phi^{-}\rangle_{A_{2}2}\right\} |1\rangle_{3}\right.\\
 & + & \left.\left\{ |\psi^{+}\rangle_{A_{1}1}|\phi^{+}\rangle_{A_{2}2}-|\psi^{-}\rangle_{A_{1}1}|\phi^{-}\rangle_{A_{2}2}+|\phi^{+}\rangle_{A_{1}1}|\psi^{+}\rangle_{A_{2}2}-|\phi^{-}\rangle_{A_{1}1}|\psi^{-}\rangle_{A_{2}2}\right\} |0\rangle_{3}\right);
\end{array}\label{eq:-1-1}
\end{equation}
and if Charlie's measurement result is $|1\rangle$ and Alice encodes
$0$, the state is
\begin{equation}
\begin{array}{lcl}
|\psi^{\prime}\rangle & = & \frac{1}{2\sqrt{2}}\left(\left\{ |\psi^{+}\rangle_{A_{1}1}|\phi^{+}\rangle_{A_{2}2}+|\psi^{-}\rangle_{A_{1}1}|\phi^{-}\rangle_{A_{2}2}+|\phi^{+}\rangle_{A_{1}1}|\psi^{+}\rangle_{A_{2}2}+|\phi^{-}\rangle_{A_{1}1}|\psi^{-}\rangle_{A_{2}2}\right\} |0\rangle_{3}\right.\\
 & + & \left.\left\{ |\psi^{+}\rangle_{A_{1}1}|\psi^{+}\rangle_{A_{2}2}+|\psi^{-}\rangle_{A_{1}1}|\psi^{-}\rangle_{A_{2}2}+|\phi^{+}\rangle_{A_{1}1}|\phi^{+}\rangle_{A_{2}2}+|\phi^{-}\rangle_{A_{1}1}|\phi^{-}\rangle_{A_{2}2}\right\} |1\rangle_{3}\right).
\end{array}\label{eq:-3-1}
\end{equation}
Thus, if Charlie's measurement outcome on qubit 4 and combined parity
of Alice's Bell measurements on $\left(A_{1},1\right)$ and $\left(A_{2},2\right)$
are same (opposite) then Bob knows her measurement outcome on qubit
3 is same (opposite) as Alice's encoding.}
\end{description}
Notice that message encoded qubits \textcolor{black}{$A_{1}$ and
$A_{2}$ are prepared and encoded locally by Alice and are not accessible
to Eve. This will be clear from the optical design of corresponding
scheme in Figure \ref{fig:CDSQC}. }

\subsubsection*{\textcolor{black}{Scheme E4: }Optical design for CDSQC protocol with
entanglement swapping using single photons (polarization qubits)}

The optical implementation of Scheme \textbf{T4} works as follows.
\begin{description}
\item [{D-O\_1\textcolor{black}{:}}] \textcolor{black}{Same as }\textbf{\textcolor{black}{Ce-O\_1}}\textcolor{black}{.
The four qubit state can be written as 
\begin{equation}
|\psi\rangle=\frac{1}{2}\left(|+\rangle_{2^{'}}\left(|+\rangle_{3^{'}}|\psi^{+}\rangle_{14}+|-\rangle|\phi^{+}\rangle_{14}\right)+|-\rangle_{2^{'}}\left(|+\rangle_{3^{'}}|\phi^{+}\rangle_{14}+|-\rangle|\psi^{+}\rangle_{14}\right)\right).\label{eq:ent-sw}
\end{equation}
}
\item [{\textcolor{black}{D-O\_2:}}] \textcolor{black}{Charlie sends corresponding
photons $3^{'}$ after passing through HWP (at $2\theta=45\lyxmathsym{\textdegree}$)
to Bob and photons $1$ and $4$ to Alice without any operation.}
\item [{\textcolor{black}{D-O\_3:}}] \textcolor{black}{Bob randomly selects
the photons by using BS from the received photons to check the eavesdropping
and measures the photons in computational or diagonal basis by using
single photon detectors. Same will be happen from Alice's side, but,
Alice's photons will pass through two optical switches OS1 and OS2
to choose the decoy qubits. After that, she chooses a set of qubits
(corresponding to computational basis measurement by Bob) using optical
switches OS3 and OS4 to measure the received photons in Bell basis,
while she performs single-qubit measurements on the rest of the qubits
(corresponding to diagonal basis measurement by Bob) to check eavesdropping.}
\item [{\textcolor{black}{D-O\_4:}}] \textcolor{black}{Alice prepares entangled
state $|\psi^{+}\rangle_{A_{1}A_{2}}$ to encode her secret information.
Specifically, she applies a NOT gate using PM on one of the qubits
of the Bell state to encode ``1'' and does nothing to send ``0''.
Therefore, the combined state of Alice and Bob before her encoding
is
\[
\begin{array}{lcl}
|\psi^{\prime}\rangle & = & \frac{1}{2\sqrt{2}}\left(|\psi^{+}\rangle_{A_{1}A_{2}}|+\rangle_{2^{'}}\left(|H\rangle_{3^{'}}|\psi^{+}\rangle_{14}+|V\rangle|\phi^{+}\rangle_{14}\right)\right.\\
 & + & \left.|\psi^{+}\rangle_{A_{1}A_{2}}|-\rangle_{2^{'}}\left(|H\rangle_{3^{'}}|\phi^{+}\rangle_{14}+|V\rangle|\psi^{+}\rangle_{14}\right)\right).
\end{array}
\]
}
\item [{\textcolor{black}{D-O\_5:}}] \textcolor{black}{Alice measures qubits
$A_{1}$ and $1$ as well as $A_{2}$ and $2$ in Bell basis, while
Bob can measure his qubits in the computational basis by using single
photon detector. Subsequently, she announces her measurement outcomes,
which should reveal her message to Bob. Finally, Charlie measures
his qubit in diagonal basis and announces that when he wishes the
task accomplished.}
\end{description}

\subsection{\textcolor{black}{Quantum Dialogue}}

\textcolor{black}{QD is a two-party scheme, where Alice and Bob as
two parties wish to communicate their secret messages simultaneously
to each other. QD can be reduced from the CQD as shown in \cite{thapliyal2017quantum}.
Therefore, we have presented the feasibility of QD with single photons
and entangled photons. Here, we briefly discuss the changes to be
made in the CQD scheme to obtain the corresponding QD scheme.}

\subsubsection{\textcolor{black}{Scheme T5: QD with single photons}}

\begin{figure}
\begin{centering}
\includegraphics[scale=0.43]{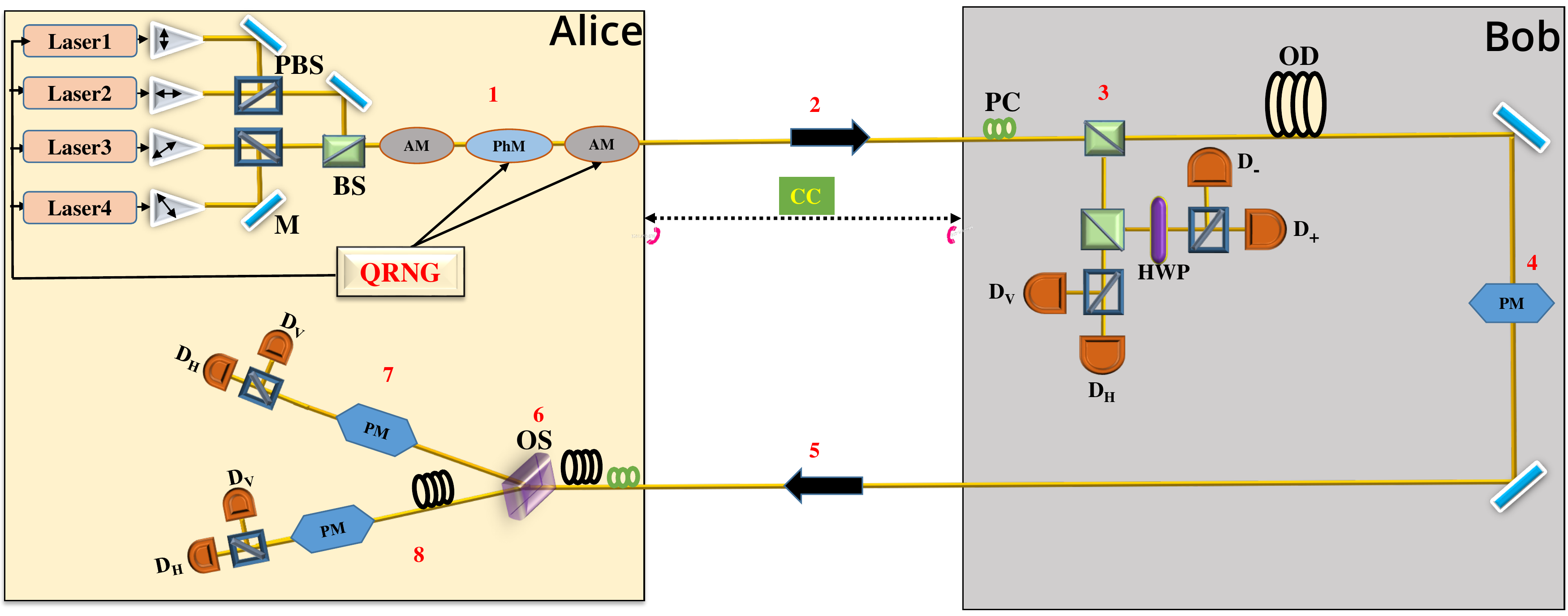}
\par\end{centering}
\centering{}\caption{\label{fig:QD}The proposed optical diagram using linear optics of
the QD scheme which is based on single photons (polarization qubits).}
\end{figure}

\textcolor{black}{QD protocol, which can be reduced from Scheme }\textbf{\textcolor{black}{T1}}\textcolor{black}{,
is summarized in the following steps:}
\begin{description}
\item [{\textcolor{black}{Q\_1:}}] \textcolor{black}{Same as }\textbf{\textcolor{black}{Cs\_1}}\textcolor{black}{,
but here Alice prepares the string.}
\item [{\textcolor{black}{Q\_2:}}] \textcolor{black}{Alice sends the string
to Bob as in }\textbf{\textcolor{black}{Cs\_2}}\textcolor{black}{. }
\item [{\textcolor{black}{Q\_3:}}] \textcolor{black}{Same as }\textbf{\textcolor{black}{Cs\_3}}\textcolor{black}{,
but Alice and Bob perform eavesdropping checking.}
\item [{\textcolor{black}{Q\_4-7:}}] \textcolor{black}{Same as }\textbf{\textcolor{black}{Cs\_4}}\textcolor{black}{-}\textbf{\textcolor{black}{7}}\textcolor{black}{,
but roles of Alice and Bob are reversed from Scheme }\textbf{\textcolor{black}{T1}}\textcolor{black}{.}
\item [{\textcolor{black}{Q\_8:}}] \textcolor{black}{Same as}\textbf{\textcolor{black}{{}
Cs\_8}}\textcolor{black}{, while Alice prepares the initial string,
so she knows the basis used for its preparation. Therefore, in the
end, Alice announces both initially prepared state and final states
on measurement. Without loss of generality the initial state can be
assumed public knowledge.}
\end{description}
\textcolor{black}{The above summarized QD protocol using single photons
can be realized by optical elements and the optical circuit for that
is illustrated in Figure \ref{fig:QD}, and the same is explained
below in steps.}

\subsubsection*{\textcolor{black}{Scheme E5: }Optical design for QD protocol using
single photons (polarization qubits)}

Scheme \textbf{T4} can be implemented in the following steps:
\begin{description}
\item [{Q-O\_1:}] \textcolor{black}{Same as }\textbf{\textcolor{black}{Cs-O\_1}}\textcolor{black}{,
but here Alice prepares the string.}
\item [{\textcolor{black}{Q-O\_2:}}] \textcolor{black}{Alice sends it to
Bob. }
\item [{\textcolor{black}{Q-O\_3-7:}}] \textcolor{black}{Same as }\textbf{\textcolor{black}{Cs-O\_3}}\textcolor{black}{-}\textbf{\textcolor{black}{7}}\textcolor{black}{.}
\item [{\textcolor{black}{Q-O\_8:}}] \textcolor{black}{Same as}\textbf{\textcolor{black}{{}
Cs-O\_8}}\textcolor{black}{, while Alice prepares the initial string
so she knows the basis of string.}
\end{description}
\textcolor{black}{Similarly, the rest of the QD schemes using single
photon (from }Scheme \textbf{T2}\textcolor{black}{) and entangled
s}tates \textcolor{black}{(from }Scheme \textbf{T3}\textcolor{black}{)
}can be reduced from corresponding CQD schemes. Here, we have avoided
such repetition and mentioned only briefly for the sake of completeness.

\subsection{\textcolor{black}{Quantum secure direct communication/Direct secure
quantum communication}}

\textcolor{black}{In quantum secure direct communication scheme, messages
are transmitted directly in a deterministic and secure manner from
Alice to Bob. A QSDC scheme can be viewed as a quantum dialogue, but
the difference is only here that one party is restricted to encode
the identity only. Similarly, a DSQC scheme can be deduced from a
QSDC or CDSQC scheme where the receiver does not encode his/her message
and requires an additional 1 bit of classical communication from the
sender to decode the message.}

\subsection{\textcolor{black}{Quantum key agreement}}

\textcolor{black}{The proposed optical designs can also be used to
reduce QKA schemes. Specifically, in quantum key agreement, all parties
take part in the key generation process and none can control the key
solely. A QKA scheme has been proposed in the recent past using a
modified version of QSDC/DSQC scheme \cite{shukla2014protocols} which
can be realized experimentally using present optical designs. Precisely,
using QSDC scheme, one party sends his/her raw key to another party
in a secure manner, while the other party publicly announces his/her
key and the final key is combined from both raw keys. Therefore, the
optical designs can be used for secure transmission of the first party's
raw key.}

\subsection{\textcolor{black}{Quantum key distribution}}

\textcolor{black}{The optical designs can be used to describe prepare-and-measure
QKD schemes, too. Specifically, decoy qubit based QKD \cite{lo2005decoy,rosenberg2007long}
can be visualized from }\textbf{\textcolor{black}{Cs-O\_1-3 }}\textcolor{black}{if
Charlie (Bob) is the sender (receiver) and they share a quantum key
by the end of this scheme. Similarly, the entangled state based CQD
can be used to describe BBM scheme \cite{bennett1992quantumcryptography}.}

\section{\textcolor{black}{Conclusions \label{sec:Conclusions}}}

\textcolor{black}{We have provided optical circuits for various quantum
cryptographic schemes not yet experimentally realized (CQD, CDSQC
with entanglement swapping, QD, etc.) with single and entangled photons
(GHZ-like) as well as modified version of three-stage protocol. Interestingly,
}most of the \textcolor{black}{designed optical circuits can be realized
using both optical fiber based and open air based architectures. Specifically,
we have shown feasibility (by providing optically implementable modified
schemes) of three-party two-way scheme, i.e., a CQD scheme using single
photons, a CQD scheme using tripartite entanglement, and a five-stage
CQD scheme using single photons. We further deduced optical designs
for three-party one-way and two-party one- and two-way quantum cryptographic
schemes from these three CQD scheme, namely CDSQC, QD, QSDC, QKA,
QKD. We have also proposed an entanglement swapping based DSQC and
CDSQC schemes where encoded photons are both prepared and measured
locally in sender's lab and are thus never accessible to the eavesdropper.
Additionally, the present results establish feasibility of several
other secure communication schemes, such as asymmetric QD \cite{banerjee2017asymmetric}
and multiparty counterpart \cite{banerjee2018quantum} of these schemes.
Further, quantum solutions for some socio-economic solutions (which
can be viewed as secure multiparty computation tasks) based on secure
direct quantum communication schemes, for instance, sealed-bid auction
\cite{sharma2017quantum}, voting \cite{thapliyal2017protocols},
e-commerce \cite{thapliyal2019quantum}, private comparison \cite{thapliyal2018orthogonal}.
Similar studies can be extended to semiquantum \cite{shukla2017semi}
and continuous variable \cite{saxena2019continuous,srikara2019CQD}
counterparts of direct secure quantum communication schemes.}

\textcolor{black}{Being a theoretical physics group, we could not
realize these optical circuits in the laboratory, but others having
laboratory facility are expected to find the work reported here as
interesting enough to perform experiments to realize the optical circuits
designed in this paper. Further, we believe that this rigorous study
will enable others to design new schemes for quantum communication
which will be experimentally realizable and to experimentally realize
such new schemes as well as the existing schemes (after suitable modification
following the approach adopted here) of various quantum communication
tasks. }

\section*{Acknowledgment}

AP acknowledges the support from Interdisciplinary Cyber Physical
Systems (ICPS) programme of the Department of Science and Technology
(DST), India, Grant No.: DST/ICPS/QuST/Theme-1/2019/14. KT acknowledges
the financial support from the Operational Programme Research, Development
and Education - European Regional Development Fund project no. CZ.02.1.01/0.0/0.0/16\_019/0000754
of the Ministry of Education, Youth and Sports of the Czech Republic.

\end{document}